# Optimization of PI Coefficients in DSTATCOM Nonlinear Controller for Regulating DC Voltage using Particle Swarm Optimization

## Dr. Ramesh Kumar[1], Dilawar Hussain[2], Ruchita[3]

[1]*Associate Professor, Dept. of Electrical Engineering,*
*National Institute Of Technology Patna, India*
*rknitpatna@gmail.com*
[2]*M.Tech, Dept. of Electrical Engineering,*
*National Institute Of Technology Patna, India*
*dilawarnitp@gmail.com*
[3]*B.Tech (IVth year), Department of Electronics & Communication Engineering,*
*Rajasthan Institute of Engineering & Technology Jaipur, Rajasthan, India*
*ruchitahoney91@gmail.com*

**Abstract**-*Non-linear controller is preferred to linear controller due to non-linear operation of DSTATCOM. System dynamic can be improved by regulating and fixing the capacitor DC voltage in DSTATCOM. The nonlinear control is based on exact linearization via feedback. There is a PI controller in this system to regulate DC voltage. In conventional scheme, the trial and error method is used to determine PI values. Exact calculation to optimize PI coefficients can be carried out to reduce disturbances in DC link voltage and thus, in this paper, Particle Swarm Optimization is applied. As a result, Capacitor voltage tracks the reference values which have less vibration than conventional status. Both trial and error method and PSO are implemented. A set of corresponding diagrams achieved by these two methods are offered to demonstrate the effectiveness of new method. Optimizations and Simulations are worked out in MATLAB environment.*

## 1. INTRODUCTION

The concept of static compensator (STATCOM) is based on shunt Flexible AC Transmission system (FACTS) device[1] that can regulate line voltage at the Point of Common Coupling (PCC), balance loads or compensate load reactive power by producing the desired amplitude and phase of inverter output voltage which is connected to a DC capacitor (energy storage device) . The STATCOM connected in distributed system is called as DSTATCOM. There are many possible configurations of Voltage Source Inverter (VSI) and consequently many different configurations of DSTATCOM are available. There are many different strategies such as Proportional Integral Controller, Sliding mode Controller and Nonlinear controller to control DSTATCOM [2-3]. Because of nonlinear operation of DSTATCOM, Nonlinear controller will be preferred in comparison with linear method. System dynamic can be improved by reducing and fixing the capacitor DC voltage in DSTATCOM. The nonlinear control is based on exact linearization via feedback [4]. There is a PI controller in this control system to regulate DC voltage. Moreover, few chosen sets of PI parameters may not be suitable for all ranges of operating points and finding these values are time consuming and complex. Vector-controlled techniques [5] are used to control the converter's currents leading to fast reactive current control. In nonlinear controller, the Generalized Averaged Method [6] has been used to determine the nonlinear time invariant continuous model of the system. This technique is particularly interesting because it transforms a nonlinear system into a linear one [7] in terms of its input-output relationship. Only q-axis current will regulate, but it should be noted that unlike other shunt compensators, large energy storage device that have almost constant voltage, makes DSTATCOM more robust and it also enhances the response speed. Therefore, there are control objectives implemented in DSTATCOM. First is q-axes current and the second is capacitor voltage in DC link. Fortunately, q-axes current tracks its corresponding reference value perfectly, but capacitor voltage is not fixed on reference ideally because of presence of a proportional-integral controller between the reference of the d-axes current and DC link voltage error. In other words, the performance indices (settling time, rise time and over shoot) are notable values. Thus, the optimized and exact determination of PI gains can lead to reduction in disturbances of system response.

In this process, one of the powerful and famous optimization algorithms named PARTICLE SWARM OPTIMIZATION (PSO) [9] is applied to find optimized values of PI gains. Particle Swarm Optimization (PSO) is a computational method that optimizes a problem iteratively trying to improve a solution with regard to a given measure of quality. This paper deals with defined objective function. The new PI coefficients, calculated in these ways, are implemented in controller to demonstrate the improvement of convergence speed, reduction of error, the overshoot in capacitor voltage and other circuit parameters. The results are compared with trial and error method.

In conventional scheme, the trial and error method is used to determine PI values. Exact calculation of optimized PI coefficients will be carried out to reduce disturbances in DC link voltage by PARTICLE SWARM OPTIMIZATION.

2. **DSTATCOM MODEL**

A simplified DSTATCOM model shown in figure (1) is considered. It consists of a voltage inverter, a capacitor and an inductance. $V_a, V_b, V_c$. $E_a, E_b, E_c$ are the line voltages and inverter output voltages respectively. $V_{dc}$ is the DC voltage.

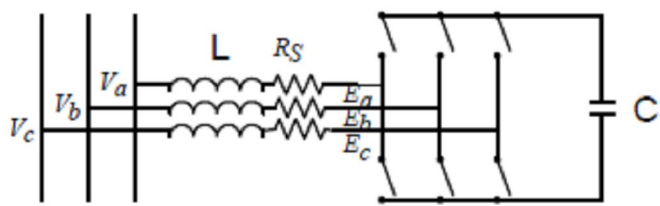

Figure1. DSTATCOM model

The ac system is described as -

$$\frac{dIa}{dt} = -\frac{Rs}{Ls}Ia + \frac{1}{Ls}(Va - Ea) \quad (1)$$

$$\frac{dIb}{dt} = -\frac{Rs}{Ls}Ib + \frac{1}{Ls}(Vb - Eb) \quad (2)$$

$$\frac{dIc}{dt} = -\frac{Rs}{Ls}Ic + \frac{1}{Ls}(Vc - Ec) \quad (3)$$

These equations can be written in matrix form as-

$$\frac{d}{dt}\begin{bmatrix}Ia\\Ib\\Ic\end{bmatrix} = -\frac{Rs}{Ls}\begin{bmatrix}1 & 0 & 0\\0 & 1 & 0\\0 & 0 & 1\end{bmatrix}\begin{bmatrix}Ia\\Ib\\Ic\end{bmatrix} + \frac{1}{Ls}(Vabc - Eabc) \quad\quad (4)$$

Also, voltage in inverter side is

$$V_{abc} = \frac{Vdc}{6}\begin{bmatrix}2 & -1 & -1\\-1 & 2 & -1\\-1 & -1 & 2\end{bmatrix}\begin{bmatrix}Ua\\Ub\\Uc\end{bmatrix} \quad\quad (5)$$

Where $U_{abc}$ is the switching vector,
$U_{abc} = [Ua\ Ub\ Uc]^T$

For DC part,

$$\frac{dV_{dc}}{dt} = \frac{I_{dc}}{C} - \frac{V_{dc}}{RC} \qquad (6)$$

$$I_{dc} = [U_a \ U_b \ U_c]^T [I_a \ I_b \ I_c] \qquad (7)$$

The angular velocity of the AC voltage or current vectors is equal to ω. Consider a system of reference (d, q) rotating [10] at the same speed, and φ the angle between d-axis and line voltage vector $\overline{E}$. The model of the ac side in this system of reference is:

$$\frac{dI_{qd}}{dt} = \begin{bmatrix} -\frac{R_s}{L_s} & -\omega \\ \omega & -\frac{R_s}{L_s} \end{bmatrix} I_{qd} + \frac{1}{L_s}(E_{qd} - V_{qd}) \qquad (8)$$

The power are expressed by equation (9)

$$P = \frac{3}{2}(E_d I_d + E_q I_q) \qquad (9)$$

$$Q = \frac{3}{2}(E_d I_q - E_q I_d) \qquad (10)$$

If, $\varphi = 0$ Then, Eq = 0
And reactive power proportional to EdIq. Now to control the reactive power, it is sufficient to control Iq. At phase angle $\varphi = 0$, Ed = E, and Eq = 0.
So, equation becomes as –

$$Q = \frac{3}{2} E_d I_q = \frac{3}{2} E I_q \qquad (11)$$

For capacitor voltage,

$$P = V_{dc} C \frac{dV_{dc}}{dt} \qquad (12)$$

Now, there are two possible kinds of solution for controlling the DSTATCOM as –

**(i) Variable structure control such as sliding mode control:** In this case, the previous model can be used to verify the sliding condition. The switching frequency is high. It avoids important communication losses.

**(ii) Continuous control associated to full wave communication of the inverter:** The control variable α is the firing angle with reference to the network voltage Ej zero crossing. A generalized averaging method [6] is used to get a continuous time invariant model of the converter. The averaged equations are

$$\frac{dI_d}{dt} = \frac{R_s}{L_s} I_d + W I_q - \frac{M \cos\alpha}{L_s} V_{dc} + \frac{E}{L_s} \qquad (13)$$

$$\frac{dI_q}{dt} = -\frac{R_s}{L_s} I_q - W I_d - \frac{M \sin\alpha}{L_s} V_{dc} \qquad (14)$$

$$\frac{dV_{dc}}{dt} = \frac{M \cos\alpha}{C} I_d + \frac{M \sin\alpha}{C} I_q \qquad (15)$$

This nonlinear model can be written as in matrix form,

$$\frac{d}{dt}\begin{pmatrix} Id \\ Iq \\ Vdc \end{pmatrix} = \begin{pmatrix} -\frac{Rs}{Ls} & \omega & -\frac{M\cos\alpha}{Ls} \\ -\omega & -\frac{Rs}{Ls} & -\frac{M\sin\alpha}{Ls} \\ \frac{M\cos\alpha}{c} & \frac{M\sin\alpha}{c} & 0 \end{pmatrix} \begin{pmatrix} Id \\ Iq \\ Vdc \end{pmatrix} + \frac{1}{Ls}\begin{pmatrix} Vs \\ 0 \\ 0 \end{pmatrix} \quad \ldots\ldots (16)$$

### 3. NONLINEAR CONTROL SCHEME FOR DSTATCOM

According to the Non-linear control law:-

$$\dot{x} = f(x) + \sum_{i=1}^{m} g_i(x) u_i$$

$$y_i = h_i(x)$$

$$L_f h(x) = \frac{\partial h(x)}{\partial x} f(x)$$

$$[L_{g1} L_f^k h_i(x) \quad L_{g2} L_f^k h_i(x) \quad \cdots \quad L_{gm} L_f^k h_i(x)] = 0$$

$$L_{gi} L_f^{r_i - 1} h_i(x) \neq 0$$

For at Least one $1 \leq j \leq m$ ………………….. (14)

In DSTATCOM system, because of compensating the reactive power and eliminating the undesired internal dynamic, Q and Vdc are chosen as output control variables. Consequently, the modulation index (m) and firing angle α are chosen as two control input variables. (A multi input multi output system is obtained)

The above equations describe a relative degree of r = {1 1}, and a fairly standard form. By solving the problem of reproducing a reference output, the following control law can be obtained:

$$\dot{X} = f(x) + g_1(x)u_1 + g_2(x)u_2 =$$

$$\begin{bmatrix} -\frac{R_S}{L_S}x_1 & \omega x_2 & \frac{V_S}{L_S} \\ -\omega x_1 & -\frac{R_S}{L_S}x_2 & 0 \\ 0 & 0 & 0 \end{bmatrix} + \begin{bmatrix} -\frac{x_3}{L_S} \\ 0 \\ -\frac{x_1}{c} \end{bmatrix} u_1 + \begin{bmatrix} 0 \\ -\frac{x_3}{L_S} \\ -\frac{x_2}{c} \end{bmatrix} u_2$$

$$Y = \begin{bmatrix} h_1(X) \\ h_2(X) \end{bmatrix} = \begin{bmatrix} x_1 \\ x_2 \end{bmatrix}$$

Where

$$X = \begin{bmatrix} x_1 \\ x_2 \\ x_3 \end{bmatrix} = \begin{bmatrix} I_d \\ I_q \\ V_{dc} \end{bmatrix} \text{ is state vector}$$

$$U = \begin{bmatrix} u_1 \\ u_2 \end{bmatrix} = \begin{bmatrix} M\cos\alpha \\ M\sin\alpha \end{bmatrix} \text{ is control vector} \quad \ldots\ldots\ldots\ldots\ldots\ldots (14)$$

The following control low can be obtained:

$$U(t) = \begin{bmatrix} -\frac{L_S}{x_3} & 0 \\ 0 & -\frac{L_S}{x_3} \end{bmatrix} \left( \begin{bmatrix} -\frac{R_S}{L_S}x_1 + \omega x_2 + \frac{V_S}{L_S} \\ -\omega x_1 - \frac{R_S}{L_S}x_2 \end{bmatrix} + \begin{bmatrix} v_1 \\ v_2 \end{bmatrix} \right) = \begin{bmatrix} -\frac{L_S}{x_3}(v_1 + \frac{R_S}{L_S}x_1 + \omega x_2 - \frac{V_S}{L_S}) \\ -\frac{L_S}{x_3}(v_2 + \omega x_1 + \frac{R_S}{L_S}x_2) \end{bmatrix}$$

…………..(15)

Where $v_1$ and $v_2$ are output references and $v_1, v_2$ are the new inputs and $y_1, y_2$ are their corresponding outputs.

Two proportional controllers are chosen to construct the new inputs ($v_1$ and $v_2$) and an external PI controller is chosen to regulate dc link voltage (Fig.2), the system with Nonlinear control law and three controllers is described.

Considering the $I_q$ channel, the equivalent close-loop transfer function can be expressed as:

$$\frac{I_q}{I_q^*} = \frac{1}{1 + s/\lambda} \quad (16)$$

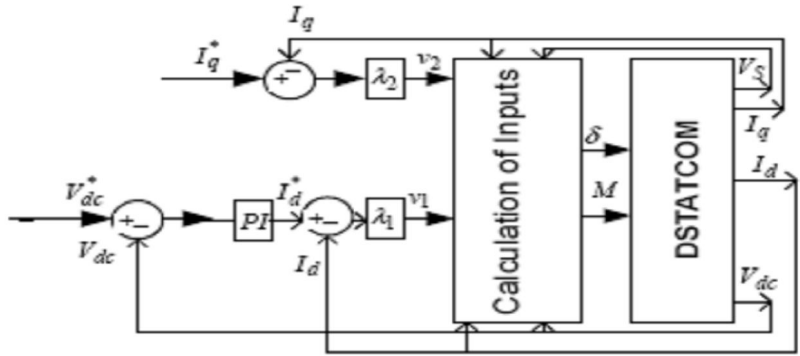

Figure2. STATCOM Controller

Where λ determines the response speed of reactive current.
Substituting Equation (15) in Equation (14) and considering equation (16), the following Equations are obtained:-

$$\begin{bmatrix} \dot{x}_1 \\ \dot{x}_2 \end{bmatrix} = \begin{bmatrix} v_1 \\ v_2 \end{bmatrix} = \begin{bmatrix} \lambda_1(x_1^* - x_1) \\ \lambda_2(x_2^* - x_2) \end{bmatrix} \quad (17)$$

By solving this differential equation, xi tends to its reference $(x_i^*)$.

4. **PARTICLE SWARM OPTIMIZATION**

PSO optimizes a problem having a population of the solutions, here dubbed particles, and moving these particles around in the search-space according to simple mathematical formulae over the particle's position and velocity. Each particle's movement is influenced by its local best known position and it is also guided towards the best known positions in the search-space, which are updated as better positions are found by other particles. This is expected to move the swarm toward the best solutions. It makes few or no assumptions about the problem being optimized and can search very large space for the solution.

The following procedure can be used for implementing the PSO algorithm.
1) Initialize the swarm by assigning a random position in the problem hyperspace to each particle.
2) Evaluate the fitness function for each particle.
3) For each individual particle, compare the particle's fitness value with its $P_{best}$. If the current value is better than the $P_{best}$ value, then set this value as the $P_{best}$ and the current particle's position, $X_i$ as $P_i$.
4) Identify the particle that has the best fitness value. The value of its fitness function is identified as $g_{best}$ and its position as $P_g$.
5) Update the velocities and positions of all the particles using (1) and (2).

Repeat steps 2–5 until a stopping criterion is met (e.g., maximum number of iterations or a sufficiently good fitness value).

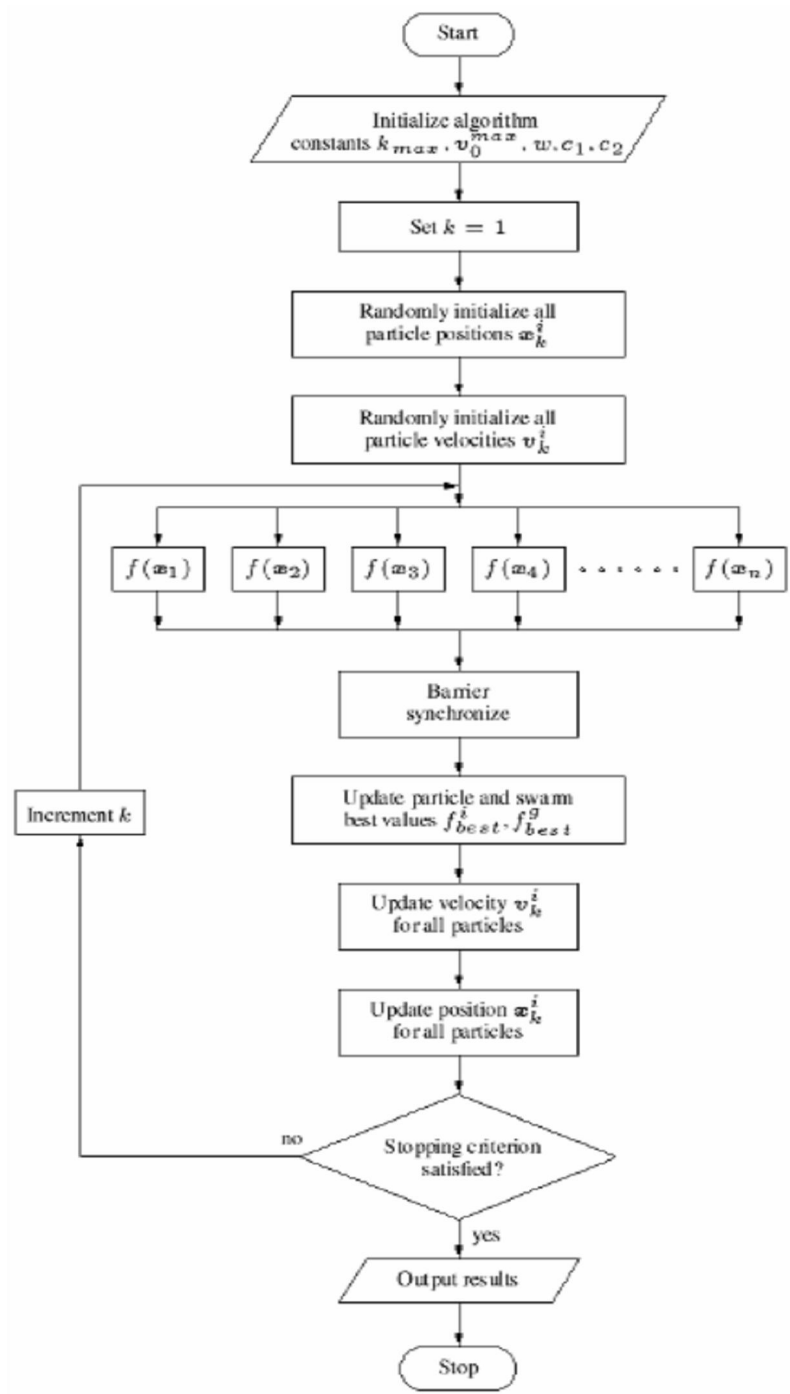

Figure 3. PSO Algorithm

5. **OBJECTIVE FUNCTION**

For best response, the essential function of a feed back control system is to reduce the error '*e(t)*' between any variable and its demanded value to zero as quickly as possible. There are four basic criteria which are commonly used-

$$\text{Integral of absolute error (IAE)} = \int_{0}^{\infty} |e(t)|.dt$$

$$\text{Integral of squared error (ISE)} = \int_0^\infty \{e(t)\}^2 \, dt$$

$$\text{Integral of time multiplied by absolute error (ITAE)} = \int_0^\infty t|e(t)| \, dt$$

$$\text{Integral of time multiplied by squared error (ISE)} = \int_0^\infty t\{e(t)\}^2 \, dt$$

For any of the above possible criteria, the best response corresponds to the minimum value of the chosen criterion. In all cases, it is either the absolute error or the square error involved. Straight forward integration of the error would produce zero result even if the system response is a constant amplitude oscillation. IAE is often used where digital simulation of a system is being employed, but it is not applicable for analytical work, because the absolute value of an error function is not generally analytic in form. The ITAE and ITSE have an additional time multiplier of the error function, which emphasises long-duration errors, and therefore these criteria are most often applied in systems requiring a fast settling time. The ITAE performance index has the advantage of producing smaller overshoots and oscillations than the IAE or ISE performance index. It is most sensitive among all. So ITAE performance index is identified as most suitable.

ITSE index is somewhat less sensitive and is not comfortable computationally, since it is not practical to integrate up to infinity.

For the DSTATCOM system, the adopted objective function is:

$$Q_f(Z) = \sum_i m_i f_i(Z)$$

Where

$$f_i(Z) = \sum_j \omega_j \int_0^T t|e_j(t)| \quad (18)$$

$f_i$ is a performance index corresponding to the $No.i$ objective. $m_i$ is a weighted factor corresponding to the objective. $e_j(t)$ is the error between the real value of the $No.j$ controlled variable and its desired value. $\omega_i$ is the weighted factor corresponding to the $No.j$ controlled variable. Vector $Z = [Z_1, Z_2, \ldots, Z_n]$ is the control system parameters, i.e. PI parameters.

For the DSTATCOM, the objective Function deduced is expressed as:

$$f(z) = 1000 \int_0^T t|Vdc - Vdcreff| \, dt \quad (19)$$

Where,

$$Z = [K_P \quad K_I]$$

The equation is used when Vdc is regulating.

6. **SIMULATION RESULTS**

The case study parameters of distribution system and DSTATCOM are as follows:
  C=4900 (μF),
  F=50 Hz,
  Rs=.28 (Ω),
  L= 0.0013 (H),
  Va=110rms(L-L) (V),

Vdc=200 (V),
Initial voltage= 200 (V)
The reference $I_q$ has a step change from zero to 15 (A) at t=0.02S. λ1 and λ2 are selected equal to 1000.

First of all, the effect of PI gains on voltage regulation is shown in the following figures by offering random PI gains, then by computed through trial and error method and finally PSO is implemented for the operation of DSTATCOM.

**Voltage regulation result**

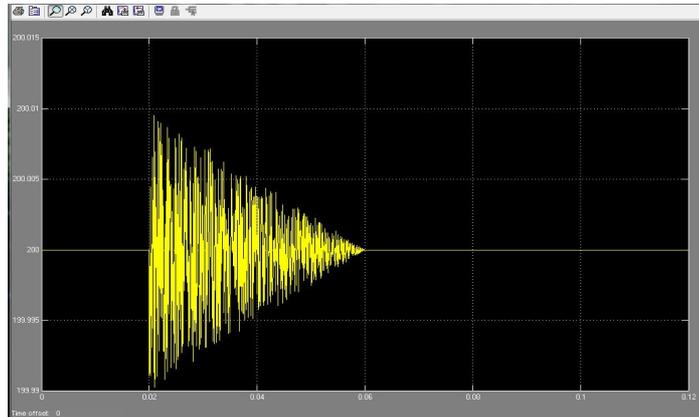

Fig.4. scope result (when gains Kp=3.214 and Kp=14.245) At random value.

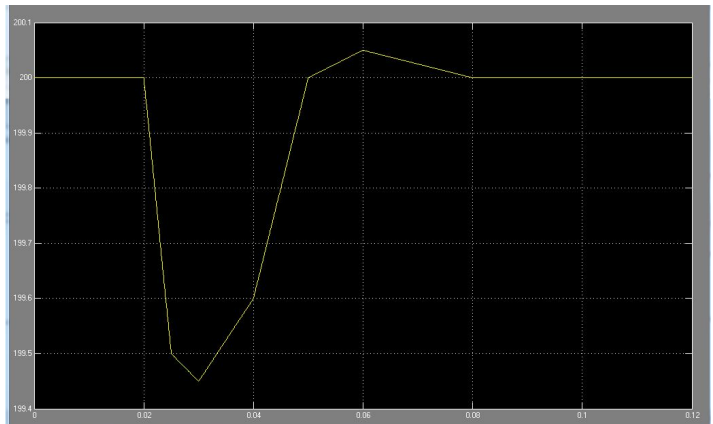

Fig.5. Scope result (when gains Kp=1 and Ki=70); By heat-and-trial value.

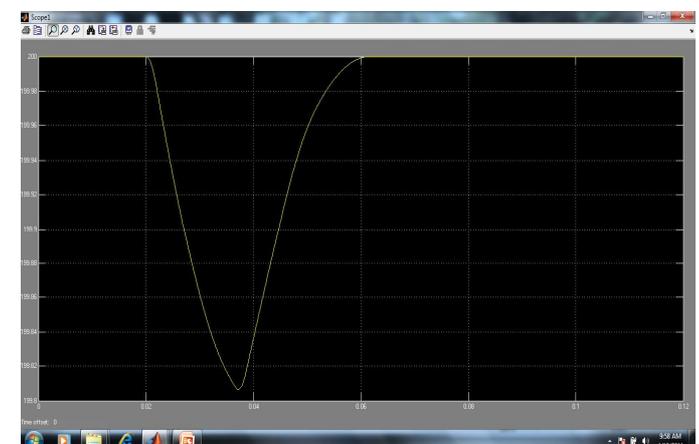

Fig.6. Scope result (when gains Kp=415.2451 and Ki=31.0245); By PSO value.

**d-axis current result**

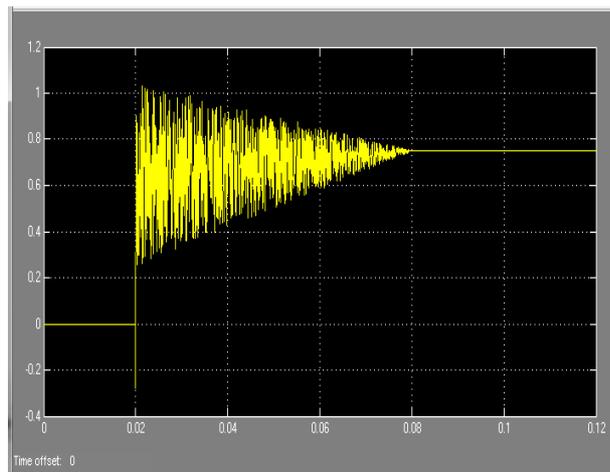
Fig.7. Scope result (When Kp=3.2145 and Ki= 14.2455); By random value.

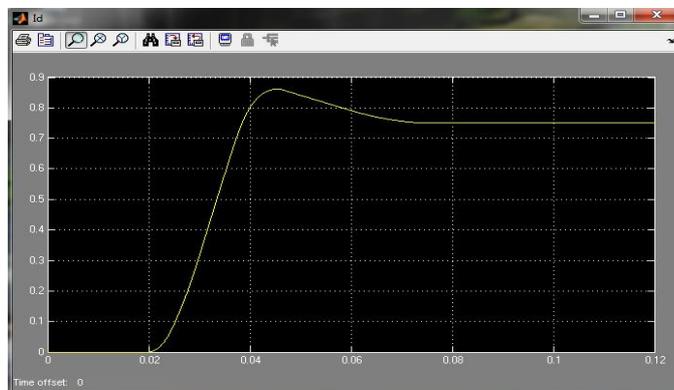
Fig.8. Scope result (When Kp=1 and Ki=70); By heat-and-trail value.

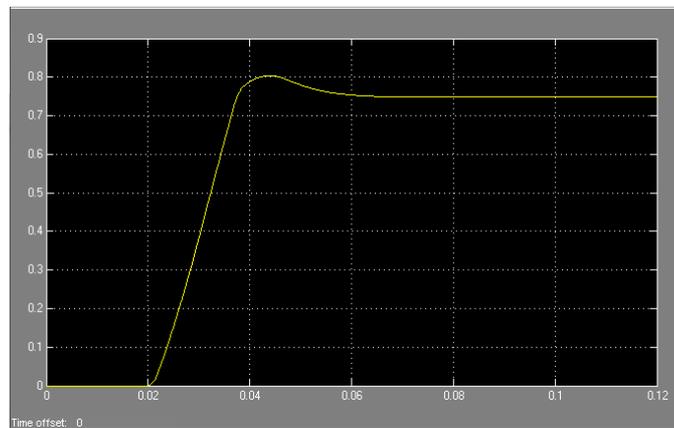
Fig.9. Scope result (When Kp=415.2145 and Ki=31.0245); By PSO value.

## 7. CONCLUSION

The nonlinear control method of the DSTATCOM which is based on the exact linearization via feedback uses one proportional–integral controller. The PI values have a remarkable influence on responses of system variables such as line current and DC link voltage. Normal way to calculate these coefficients is trial and error method. Particle Swarm Optimization method with objective function leads to a better regulation of DC link voltage, d and q axis currents and other circuit parameters. Actually, the time of reaching to steady state value, settling time the fluctuations and overshoot decrease.

Future work for the current system is to design and develop algorithm for other controller.

## 8. REFERENCES

[1] Hingorani, N. G.; and Gyugyi, L.; (1999) "Understanding FACTS, Concepts and Technology of Fluxible AC Transmission Systems." 0-7803-3455-8, New York

[2] S. Iyer, A. Ghosh and A.Joshi," Inverter topologies for DSTATCOM applications-a simulation study," Electr. Power syst. Res.75(2005), pp. 161-170 (Elsevier)

[3] Lauttamus, P.; Tuusa, H.; "Comparison of Five-Level Voltage-Source Inverter Based STATCOMs" IEEE Power Conversion Conference, 2-5 April 2007 Page(s):659 – 666

[4] Petitclair, P.; Bacha, S.; Ferrieux, J. P.; (1997) "Optimized linearization via feedback control law for a STATCOM" Pages 880 – 885, France

[5] Schauder, C.; Mehta, H.; (1993) "Vector analysis and control of advanced static VAR compensators" IEE Proceeding on Generation, Transmission and Distribution Volume 140, Pages:299-306

[6] Sanders, S.R.; Noworolski, J.M.; Liu, X.Z.; Verghese, G.C.;(1991) "Generalized averaging method for power conversion circuits" IEEE Transaction on Power Electronics, Vol. 6, Page(s):251-259

[7] Soto, D.; Pena, R.; (2004) "Nonlinear control strategies for cascaded multilevel STATCOMs" IEEE Transaction on Power Electronics, Volume 19,Page(s):1919 – 1927

[8] Yazdani, A.; Crow, M.L.; Guo, J.;(2008) "A comparison of linear and nonlinear STATCOM control for power quality enhancement" IEEE Power and Energy Society General Meeting - Conversion and Delivery of Electrical Energy, Page(s):1 – 6

[9] Valle, Yamille del.; Venayagamoorthy, Ganesh Kumar.; Mohagheghi, Salman.; Hernandez, Jean-Carlos.; Harley, Ronald G.; (2008) "Particle swarm optimization: Basic concepts, Variants and Applications in Power systems" pages 171-195,

[10] Bhimbra, P.S.; Generalized theory of Electric machine Volume 2

**Authors' Profile:**

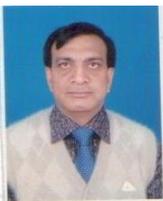
**Dr. Ramesh kumar** received the B.Sc (Engg.), Electrical, at B.C.E./NIT Patna, the M.Sc. degree in Electrical Engg. & Ph.D. from Patna University, Patna. He is now Associate Professor in Department of Electrical Engg.at NIT Patna. rknitpatna@gmail.com
Postal Address -
Associate Professor, Department Of electrical Engineering, National Institute Of Technology Patna, Patna, Bihar, India 800-005

**Dilawar Hussain** has received B.Sc. ENGGINEERING (Electronics & communication) in 2008 with first class with Distinction from Muzaffarpur Institute of Technology (MIT) Muzaffarpur, India and M.Tech degree with specialization Control Systems from Department of Electrical Engineering, National Institute of Technology (NITP), Patna, India. His field of interest includes Particle Swarm Optimization and control systems.

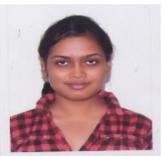
**Ruchita** is a student of Final Year, Electronics & Communication Engineering at Rajasthan Institute of Engineering & Technology Jaipur, Rajasthan, India.